\documentclass[aps,nofootinbib,twocolumn,superscriptaddress]{revtex4}%
\usepackage{hyperref}
\usepackage{amsmath}
\usepackage{amsfonts}
\usepackage{amssymb}
\usepackage{graphicx}
\usepackage{color}%
\providecommand{\U}[1]{\protect\rule{.1in}{.1in}}

\begin{document}
\title{The Exact $SL(K+3,%
\mathbb{C}
)$ Symmetry of String Scattering Amplitudes}
\author{Sheng-Hong Lai}
\email{xgcj944137@gmail.com}
\affiliation{Department of Electrophysics, National Chiao-Tung University, Hsinchu, Taiwan, R.O.C.}
\author{Jen-Chi Lee}
\email{jcclee@cc.nctu.edu.tw}
\affiliation{Department of Electrophysics, National Chiao-Tung University, Hsinchu, Taiwan, R.O.C.}
\author{Yi Yang}
\email{yiyang@mail.nctu.edu.tw}
\affiliation{Department of Electrophysics, National Chiao-Tung University, Hsinchu, Taiwan, R.O.C.}
\author{}
\date{\today }

\begin{abstract}
We discover that the $26D$ open bosonic string scattering amplitudes (SSA) of
three tachyons and one arbitrary string state can be expressed in terms of the
D-type Lauricella functions with associated $SL(K+3,%
\mathbb{C}
)$ symmetry. As a result, SSA and symmetries or relations among SSA of
different string states at various limits calculated previously can be
rederived. These include the linear relations conjectured by Gross
\cite{GM,Gross,GrossManes} and proved in \cite{ChanLee1,ChanLee2,CHL,PRL,
CHLTY,susy} in the hard scattering limit, the recurrence relations in the
Regge scattering limit \cite{KLY,LY,LY2014} and the extended recurrence
relations in the nonrelativistic scattering limit \cite{LLY} discovered
recently. Finally, as an application, we calculate a new recurrence relation
of SSA which is valid for \textit{all} energies.

\end{abstract}
\maketitle

\noindent\textit{Introduction} It has long been believed that there exist huge
hidden spacetime symmetries of string theory. As a consistent theory of
qnantum gravity, string theory contains no free parameter and an infinite
number of higher spin string states. On the other hand, the very soft
exponential fall-off behavior of string scattering amplitudes (SSA) in the
hard scattering limit, in constrast to the power law behavior of those of
quantum field theory, strongly suggests the existence of infinite number of
relations among SSA of different string states. These relations or symmetries
soften the UV structure of quantum string theory. Indeed, this kind of
infinite relations were conjectured by Gross \cite{GM,Gross,GrossManes} and
later explicitly proved in \cite{ChanLee1,ChanLee2,CHL,PRL, CHLTY,susy}, and
can be used to reduce the number of independent hard SSA from $\infty$ down to
$1$.

Historically, there were at least three approaches to probe stringy symmetries
of higher spin string states. These include the gauge symmetry of Witten
string field theory, the conjecture of Gross \cite{Gross} on symmetries or
linear relations among SSA of different string states in the hard scattering
limit \cite{GM,Gross,GrossManes} and Moore's bracket algebra approach
\cite{Moore,Moore1,CKT} of stringy symmetries. See a recent review
\cite{review} for some connections of these three approaches.

Recently, it was found that the Regge SSA of three tachyons and one arbitrary
string states can be expressed in terms of a sum of Kummer functions $U$
\cite{KLY,LY,LY2014}, which soon later were shown to be the first Appell
function $F_{1}$ \cite{LY2014}. Regge stringy symmetries or recurrence
relations \cite{LY,LY2014} were then constructed and used to reduce the number
of independent\ Regge SSA from $\infty$ down to $1$. Moreover, an interesting
link between Regge SSA and hard SSA was found \cite{KLY,LYY}, and for each
mass level the ratios among hard SSA can be extracted from Regge SSA. This
result enables us to argue that the known $SL(5;C)$ dynamical symmetry of the
Appell function $F_{1}$ \cite{sl5c} is crucial to probe high energy spacetime
symmetry of string theory.

More recently, the extended recurrence relations \cite{LLY} among
nonrelativistic low energy SSA of a class of string states with different
spins and different channels were constructed by using the recurrence
relations of the Gauss hypergeometry functions with associated $SL\left(  4,%
\mathbb{C}
\right)  $ symmetry\cite{sl4c}. These extended recurrence relations generalize
and extend the field theory BCJ \cite{BCJ} relations to higher spin string states.

To further uncover the structure of stringy symmetries, in this paper we
calculate the $26D$ open bosonic SSA of three tachyons and one arbitrary
string states at \textit{arbitrary} energy. We discover that these SSA can be
expressed in terms of the D-type Lauricella functions with associated $SL(K+3,%
\mathbb{C}
)$ symmetry \cite{sl4c}. As a result, all these SSA and symmetries or
relations among SSA of different string states at various limits calculated
previously can be rederived. These include the linear relations conjectured by
Gross \cite{Gross} and proved in \cite{ChanLee1,ChanLee2,CHL,PRL, CHLTY,susy}
in the hard scattering limit, the recurrence relations in the Regge scattering
limit \cite{LY,LY2014} with associated $SL(5;C)$ symmetry and the extended
recurrence relations in the nonrelativistic scattering limit \cite{LLY} with
associated $SL(4;C)$ symmetry discovered very recently.

As a byproduct from the calculation of rederiving linear relations in the hard
scatteing limit directly from Lauricella functions, we propose an identity
which generalizes the Stirling number identity \cite{KLY,LYY} used previously
to extract ratios among hard SSA from the Appell functions in Regge SSA.
Finally, as an example, we calculate a new recurrence relation of SSA which is
valid for \textit{all} energies.

\noindent\textit{Four-point string amplitudes }We will consider SSA of three
tachyons and one arbitrary string states put at the second vertex. For the 26D
open bosonic string, the general states at mass level $M_{2}^{2}=2(N-1)$,
$N=\sum_{n,m,l>0}\left(  nr_{n}^{T}+mr_{m}^{P}+lr_{l}^{L}\right)  $ with
polarizations on the scattering plane are of the form%
\begin{equation}
\left\vert r_{n}^{T},r_{m}^{P},r_{l}^{L}\right\rangle =\prod_{n>0}\left(
\alpha_{-n}^{T}\right)  ^{r_{n}^{T}}\prod_{m>0}\left(  \alpha_{-m}^{P}\right)
^{r_{m}^{P}}\prod_{l>0}\left(  \alpha_{-l}^{L}\right)  ^{r_{l}^{L}}%
|0,k\rangle. \label{state}%
\end{equation}
In the CM frame, the kinematics are defined as%
\begin{align}
k_{1}  &  =\left(  \sqrt{M_{1}^{2}+|\vec{k_{1}}|^{2}},-|\vec{k_{1}}|,0\right)
,\\
k_{2}  &  =\left(  \sqrt{M_{2}+|\vec{k_{1}}|^{2}},+|\vec{k_{1}}|,0\right)  ,\\
k_{3}  &  =\left(  -\sqrt{M_{3}^{2}+|\vec{k_{3}}|^{2}},-|\vec{k_{3}}|\cos
\phi,-|\vec{k_{3}}|\sin\phi\right)  ,\\
k_{4}  &  =\left(  -\sqrt{M_{4}^{2}+|\vec{k_{3}}|^{2}},+|\vec{k_{3}}|\cos
\phi,+|\vec{k_{3}}|\sin\phi\right)
\end{align}
with $M_{1}^{2}=M_{3}^{2}=M_{4}^{2}=-2$ and $\phi$ is the scattering angle.
The Mandelstam variables are $s=-\left(  k_{1}+k_{2}\right)  ^{2}$,
$t=-\left(  k_{2}+k_{3}\right)  ^{2}$ and $u=-\left(  k_{1}+k_{3}\right)
^{2}$. There are three polarizations on the scattering plane%
\begin{align}
e^{T}  &  =(0,0,1),\\
e^{L}  &  =\frac{1}{M_{2}}\left(  |\vec{k_{1}}|,\sqrt{M_{2}+|\vec{k_{1}}|^{2}%
},0\right)  ,\\
e^{P}  &  =\frac{1}{M_{2}}\left(  \sqrt{M_{2}+|\vec{k_{1}}|^{2}},|\vec{k_{1}%
}|,0\right)  .
\end{align}
For later use, we define%
\begin{equation}
k_{i}^{X}\equiv e^{X}\cdot k_{i}\text{ \ for \ }X=\left(  T,P,L\right)  .
\end{equation}
Note that SSA of three tachyons and one arbitrary string state with
polarizations orthogonal to the scattering plane vanish. The $\left(
s,t\right)  $ and $\left(  t,u\right)  $ channels SSA of states in
Eq.(\ref{state}) can be calculated to be%
\begin{align}
&  A_{st}^{(r_{n}^{T},r_{m}^{P},r_{l}^{L})}\nonumber\\
&  =B\left(  -\frac{t}{2}-1,-\frac{s}{2}-1\right)  \prod_{n=1}\left[
-(n-1)!k_{3}^{T}\right]  ^{r_{n}^{T}}\nonumber\\
&  \cdot\prod_{m=1}\left[  -(m-1)!k_{3}^{P}\right]  ^{r_{m}^{P}}\prod
_{l=1}\left[  -(l-1)!k_{3}^{L}\right]  ^{r_{l}^{L}}\nonumber\\
&  \cdot F_{D}^{(K)}\left(  -\frac{t}{2}-1;R_{n}^{T},R_{m}^{P},R_{l}^{L}%
;\frac{u}{2}+2-N;\tilde{Z}_{n}^{T},\tilde{Z}_{m}^{P},\tilde{Z}_{l}^{L}\right)
,\label{st}\\
&  A_{tu}^{(r_{n}^{T},r_{m}^{P},r_{l}^{L})}\nonumber\\
&  =B\left(  -\frac{t}{2}-1,-\frac{u}{2}-1\right)  \prod_{n=1}\left[
-(n-1)!k_{3}^{T}\right]  ^{r_{n}^{T}}\nonumber\\
&  \cdot\prod_{m=1}\left[  -(m-1)!k_{3}^{P}\right]  ^{r_{m}^{P}}\prod
_{l=1}\left[  -(l-1)!k_{3}^{L}\right]  ^{r_{l}^{L}}\nonumber\\
&  \cdot F_{D}^{(K)}\left(  -\frac{t}{2}-1;R_{n}^{T},R_{m}^{P},R_{l}^{L}%
;\frac{s}{2}+2-N;Z_{n}^{T},Z_{m}^{P},Z_{l}^{L}\right)  , \label{tu2}%
\end{align}
where we have defined $R_{k}^{X}\equiv\left\{  -r_{1}^{X}\right\}  ^{1}%
,\cdots,\left\{  -r_{k}^{X}\right\}  ^{k}$ with $\left\{  a\right\}
^{n}=\underset{n}{\underbrace{a,a,\cdots,a}}$, $Z_{k}^{X}\equiv\left[
z_{1}^{X}\right]  ,\cdots,\left[  z_{k}^{X}\right]  $ with $\left[  z_{k}%
^{X}\right]  =z_{k0}^{X},\cdots,z_{k\left(  k-1\right)  }^{X}$ and $z_{k}%
^{X}=\left\vert \left(  -\frac{k_{1}^{X}}{k_{3}^{X}}\right)  ^{\frac{1}{k}%
}\right\vert $,\ $z_{kk^{\prime}}^{X}=z_{k}^{X}e^{\frac{2\pi ik^{\prime}}{k}}%
$,\ $\tilde{z}_{kk^{\prime}}^{X}\equiv1-z_{kk^{\prime}}^{X}$ for $k^{\prime
}=0,\cdots,k-1$. The integer $K$ is defined to be%
\begin{equation}
\text{ }K=\underset{\{\text{for all }r_{j}^{T}\neq0\}}{\sum_{j=1}^{n}%
j}+\underset{\{\text{for all }r_{j}^{P}\neq0\}}{\sum_{j=1}^{m}j}%
+\underset{\{\text{for all }r_{j}^{L}\neq0\}}{\sum_{j=1}^{l}j}.
\end{equation}
For a given $K$, there can be SSA with different mass level $N$. The D-type
Lauricella function $F_{D}^{(K)}$ is one of the four extensions of the Gauss
hypergeometric function to $K$ variables and is defined as%
\begin{align}
&  F_{D}^{(K)}\left(  a;b_{1},...,b_{K};c;x_{1},...,x_{K}\right) \nonumber\\
=  &  \sum_{n_{1},\cdots,n_{K}}\frac{\left(  a\right)  _{n_{1}+\cdots+n_{K}}%
}{\left(  c\right)  _{n_{1}+\cdots+n_{K}}}\frac{\left(  b_{1}\right)  _{n_{1}%
}\cdots\left(  b_{K}\right)  _{n_{K}}}{n_{1}!\cdots n_{K}!}x_{1}^{n_{1}}\cdots
x_{K}^{n_{K}}%
\end{align}
where $(a)_{n}=a\cdot\left(  a+1\right)  \cdots\left(  a+n-1\right)  $ is the
Pochhammer symbol. There was a integral representation of the Lauricella
function $F_{D}^{(K)}$ discovered by Appell and Kampe de Feriet (1926)
\cite{Appell}%
\begin{align}
&  F_{D}^{(K)}\left(  a;b_{1},...,b_{K};c;x_{1},...,x_{K}\right) \nonumber\\
&  =\frac{\Gamma(c)}{\Gamma(a)\Gamma(c-a)}\int_{0}^{1}dt\,t^{a-1}%
(1-t)^{c-a-1}\nonumber\\
&  \cdot(1-x_{1}t)^{-b_{1}}(1-x_{2}t)^{-b_{2}}...(1-x_{K}t)^{-b_{K}},
\label{Kam}%
\end{align}
which was used to calculate Eq.(\ref{st}) and Eq.(\ref{tu2}). By using the
identity of Lauricella function for $b_{i}\in Z^{-}$%
\begin{align}
&  F_{D}^{(K)}\left(  a;b_{1},...,b_{K};c;x_{1},...,x_{K}\right)
=\frac{\Gamma\left(  c\right)  \Gamma\left(  c-a-\sum b_{i}\right)  }%
{\Gamma\left(  c-a\right)  \Gamma\left(  c-\sum b_{i}\right)  }\nonumber\\
\cdot &  F_{D}^{(K)}\left(  a;b_{1},...,b_{K};1+a+\sum b_{i}-c;1-x_{1}%
,...,1-x_{K}\right)  ,
\end{align}
we can rederive the string BCJ relation \cite{LLY}%
\begin{align}
\frac{A_{st}^{(r_{n}^{T},r_{m}^{P},r_{l}^{L})}}{A_{tu}^{(r_{n}^{T},r_{m}%
^{P},r_{l}^{L})}}  &  =\frac{\Gamma\left(  -\frac{s}{2}-1\right)
\Gamma\left(  \frac{s}{2}+2\right)  }{\Gamma\left(  \frac{u}{2}+2-N\right)
\Gamma\left(  -\frac{u}{2}-1+N\right)  }\nonumber\\
&  =\frac{\sin\left(  \frac{\pi u}{2}\right)  }{\sin\left(  \frac{\pi s}%
{2}\right)  }=\frac{\sin\left(  \pi k_{2}\cdot k_{4}\right)  }{\sin\left(  \pi
k_{1}\cdot k_{2}\right)  },
\end{align}
which gives another form of the $\left(  s,t\right)  $ channel amplitude%
\begin{align}
&  A_{st}^{(r_{n}^{T},r_{m}^{P},r_{l}^{L})}\nonumber\\
&  =B\left(  -\frac{t}{2}-1,-\frac{s}{2}-1+N\right)  \prod_{n=1}\left[
-(n-1)!k_{3}^{T}\right]  ^{r_{n}^{T}}\nonumber\\
&  \cdot\prod_{m=1}\left[  -(m-1)!k_{3}^{P}\right]  ^{r_{m}^{P}}\prod
_{l=1}\left[  -(l-1)!k_{3}^{L}\right]  ^{r_{l}^{L}}\nonumber\\
&  \cdot F_{D}^{(K)}\left(  -\frac{t}{2}-1;R_{n}^{T},R_{m}^{P},R_{l}^{L}%
;\frac{s}{2}+2-N;Z_{n}^{T},Z_{m}^{P},Z_{l}^{L}\right)  . \label{tu}%
\end{align}

\noindent\textit{Regge scattering limit} The relevant kinematics in Regge
limit are%
\begin{align}
k_{1}^{T}  &  =0\text{, \ \ }k_{3}^{T}\simeq-\sqrt{-t},\\
k_{1}^{P}  &  \simeq-\frac{s}{2M_{2}}\text{,\ }k_{3}^{P}\simeq-\frac{\tilde
{t}}{2M_{2}}=-\frac{t-M_{2}^{2}-M_{3}^{2}}{2M_{2}},\\
k_{1}^{L}  &  \simeq-\frac{s}{2M_{2}}\text{, }k_{3}^{L}\simeq-\frac{\tilde
{t}^{\prime}}{2M_{2}}=-\frac{t+M_{2}^{2}-M_{3}^{2}}{2M_{2}},
\end{align}
with $\tilde{z}_{kk^{\prime}}^{T}=1$, \ $\tilde{z}_{kk^{\prime}}^{P}=1-\left(
-\frac{s}{\tilde{t}}\right)  ^{1/k}e^{\frac{i2\pi k^{\prime}}{k}}\sim s^{1/k}$
and $\tilde{z}_{kk^{\prime}}^{L}=1-\left(  -\frac{s}{\tilde{t}^{\prime}%
}\right)  ^{1/k}e^{\frac{i2\pi k^{\prime}}{k}}\sim s^{1/k}.$ In the Regge
limit, the SSA in Eq.(\ref{st}) reduces to%
\begin{align}
&  A_{st}^{(r_{n}^{T},r_{m}^{P},r_{l}^{L})}\nonumber\\
\simeq &  B\left(  -\frac{t}{2}-1,-\frac{s}{2}-1\right)  \prod_{n=1}\left[
(n-1)!\sqrt{-t}\right]  ^{r_{n}^{T}}\nonumber\\
\cdot &  \prod_{m=1}\left[  (m-1)!\frac{\tilde{t}}{2M_{2}}\right]  ^{r_{m}%
^{P}}\prod_{l=1}\left[  (l-1)!\frac{\tilde{t}^{\prime}}{2M_{2}}\right]
^{r_{l}^{L}}\nonumber\\
\cdot &  F_{1}\left(  -\frac{t}{2}-1;-q_{1},-r_{1};-\frac{s}{2};\frac
{s}{\tilde{t}},\frac{s}{\tilde{t}^{\prime}}\right)  , \label{app}%
\end{align}
where $F_{1}$ is the Appell function. Eq.(\ref{app}) agrees with the result
obtained in \cite{LY2014} previously.

\noindent\textit{Hard scattering limit} In the hard scattering limit
$e^{P}=e^{L}$ \cite{ChanLee1,ChanLee2}, we can consider only the polarization
$e^{L}$ case. The relevant kinematics are%
\begin{align}
k_{1}^{T}  &  =0\text{, \ \ }k_{3}^{T}\simeq-E\sin\phi,\\
k_{1}^{L}  &  \simeq-\frac{2p^{2}}{M_{2}}\simeq-\frac{2E^{2}}{M_{2}},\\
k_{3}^{L}  &  \simeq\frac{2E^{2}}{M_{2}}\sin^{2}\frac{\phi}{2},
\end{align}
with $\tilde{z}_{kk^{\prime}}^{T}=1$, \ $\tilde{z}_{kk^{\prime}}^{L}=1-\left(
-\frac{s}{t}\right)  ^{1/k}e^{\frac{i2\pi k^{\prime}}{k}}\sim O\left(
1\right)  .$ The SSA in Eq.(\ref{st}) reduces to%
\begin{align}
&  A_{st}^{(r_{n}^{T},r_{l}^{L})}=B\left(  -\frac{t}{2}-1,-\frac{s}%
{2}-1\right) \nonumber\\
&  \cdot\prod_{n=1}\left[  (n-1)!E\sin\phi\right]  ^{r_{n}^{T}}\prod
_{l=1}\left[  -(l-1)!\frac{2E^{2}}{M_{2}}\sin^{2}\frac{\phi}{2}\right]
^{r_{l}^{L}}\nonumber\\
&  \cdot F_{D}^{(K)}\left(  -\frac{t}{2}-1;R_{n}^{T},R_{l}^{L};\frac{u}%
{2}+2-N;\left(  1\right)  _{n},\tilde{Z}_{l}^{L}\right)  .
\end{align}
One key observation of the previous hard SSA calculation
\cite{ChanLee1,ChanLee2,CHL,PRL, CHLTY,susy} was that there was a difference
between the naive energy order and the real energy order corresponding to the
$\left(  \alpha_{-1}^{L}\right)  ^{r_{1}^{L}}$ operator in Eq.(\ref{state}).
So let's pay attention to the corresponding summation and write%
\begin{align}
&  A_{st}^{(r_{n}^{T},r_{l}^{L})}=B\left(  -\frac{t}{2}-1,-\frac{s}%
{2}-1\right) \nonumber\\
&  \cdot\prod_{n=1}\left[  (n-1)!E\sin\phi\right]  ^{r_{n}^{T}}\prod
_{l=1}\left[  -(l-1)!\frac{2E^{2}}{M_{2}}\sin^{2}\frac{\phi}{2}\right]
^{r_{l}^{L}}\nonumber\\
&  \cdot\sum_{k_{r}}\frac{\left(  -\frac{t}{2}-1\right)  _{k_{r}}}{\left(
\frac{u}{2}+2-N\right)  _{k_{r}}}\frac{\left(  -r_{1}^{L}\right)  _{k_{r}}%
}{k_{r}!}\left(  1+\frac{s}{t}\right)  ^{k_{r}}\cdot\left(  \cdots\right)
\end{align}
where we have used $\left(  a\right)  _{n+m}=\left(  a\right)  _{n}\left(
a+n\right)  _{m}$. We then propose the following formula%
\begin{align}
&  \sum_{k_{r}=0}^{r_{1}}\frac{\left(  -\frac{t}{2}-1\right)  _{k_{r}}%
}{\left(  \frac{u}{2}+2-N\right)  _{k_{r}}}\frac{\left(  -r_{1}^{L}\right)
_{k_{r}}}{k_{r}!}\left(  1+\frac{s}{t}\right)  ^{k_{r}}\nonumber\\
=  &  0\cdot\left(  \frac{tu}{s}\right)  ^{0}\!+0\cdot\left(  \frac{tu}%
{s}\right)  ^{-1}\!+\dots+0\cdot\left(  \frac{tu}{s}\right)  ^{-\left[
\frac{r_{1}^{L}+1}{2}\right]  -1}\nonumber\\
&  +C_{r_{l}^{L}}\left(  \frac{tu}{s}\right)  ^{-\left[  \frac{r_{1}^{L}+1}%
{2}\right]  }+\mathit{O}\left\{  \left(  \frac{tu}{s}\right)  ^{-\left[
\frac{r_{1}^{L}+1}{2}\right]  +1}\right\}  , \label{pro}%
\end{align}
which is a generalization of the Stirling number identity proposed in
\cite{KLY} and proved in \cite{LYY}. In Eq.(\ref{pro}), $C_{r_{l}^{L}}$ is
independent of energy $E$ and depends on $r_{l}^{L}$ and possibly scattering
angle $\phi$, and the $0$ terms correspond to the naive energy order in the
hard SSA calculation. The leading order SSA in the hard scattering limit can
then be identified%
\begin{align}
&  A_{st}^{(r_{n}^{T},r_{l}^{L})}\simeq B\left(  -\frac{t}{2}-1,-\frac{s}%
{2}-1\right) \nonumber\\
&  \cdot\prod_{n=1}\left[  (n-1)!E\sin\phi\right]  ^{r_{n}^{T}}\prod
_{l=1}\left[  -(l-1)!\frac{2E^{2}}{M_{2}}\sin^{2}\frac{\phi}{2}\right]
^{r_{l}^{L}}\nonumber\\
&  \cdot C_{r_{l}^{L}}\left(  E\sin\phi\right)  ^{-2\left[  \frac{r_{1}^{L}%
+1}{2}\right]  }\cdot\left(  \cdots\right) \nonumber\\
&  \sim E^{N-\sum_{n\geq2}nr_{n}^{T}-\left(  2\left[  \frac{r_{1}^{L}+1}%
{2}\right]  -r_{1}^{L}\right)  -\sum_{l\geq3}lr_{l}^{L}}\nonumber\\
&  \Rightarrow r_{n\geq2}^{T}=r_{l\geq3}^{L}=0\text{ and }r_{1}^{L}=2m,
\end{align}
which means for $r_{l}^{L}=1,3,5,\cdots$, the amplitudes are of subleading
order in energy. This is consistent with the previous results
\cite{ChanLee1,ChanLee2,CHL,PRL, CHLTY,susy}. We further propose that
$C_{r_{l}^{L}}=\frac{\left(  2m\right)  !}{m!}$ and is $\phi$ independent for
$r_{l}^{L}=2m$ in Eq.(\ref{pro}). We have verified Eq.(\ref{pro}) for
$r_{1}=0,1,2$\bigskip$\cdots,10$. Finally the leading order SSA in the hard
scattering limit, i.e. $r_{1}^{T}=N-2m-2$, $r_{1}^{L}=2m$ and $r_{2}^{L}=q$,
can be calculated to be%
\begin{align}
&  A_{st}^{(N-2m-2q,2m,q)}\nonumber\\
&  \simeq B\left(  -\frac{t}{2}-1,-\frac{s}{2}-1\right)  \left(  E\sin
\phi\right)  ^{N}\frac{\left(  2m\right)  !}{m!}\left(  -\frac{1}{2M_{2}%
}\right)  ^{2m+q}\nonumber\\
&  =(2m-1)!!\left(  -\frac{1}{M_{2}}\right)  ^{2m+q}\left(  \frac{1}%
{2}\right)  ^{m+q}A_{st}^{(N,0,0)}%
\end{align}
which is consistent with the previous result \cite{ChanLee1,ChanLee2,CHL,PRL,
CHLTY,susy}.

\noindent\textit{Nonrelativistic scattering limit} In this limit $|\vec{k_{1}%
}|\ll M_{2}$, we have%
\begin{align}
k_{1}^{T}  &  =0,k_{3}^{T}=-\left[  \frac{\epsilon}{2}+\frac{(M_{1}+M_{2}%
)^{2}}{4M_{1}M_{2}\epsilon}|\vec{k_{1}}|^{2}\right]  \sin\phi,\\
k_{1}^{L}  &  =-\frac{M_{1}+M_{2}}{M_{2}}|\vec{k_{1}}|+O\left(  |\vec{k_{1}%
}|^{2}\right)  ,\\
k_{3}^{L}  &  =-\frac{\epsilon}{2}\cos\phi+\frac{M_{1}+M_{2}}{2M_{2}}%
|\vec{k_{1}}|+O\left(  |\vec{k_{1}}|^{2}\right)  ,\\
k_{1}^{P}  &  =-M_{1}+O\left(  |\vec{k_{1}}|^{2}\right)  ,\\
k_{3}^{P}  &  =\frac{M_{1}+M_{2}}{2}-\frac{\epsilon}{2M_{2}}\cos\phi
|\vec{k_{1}}|+O\left(  |\vec{k_{1}}|^{2}\right)
\end{align}
where $\epsilon=\sqrt{(M_{1}+M_{2})^{2}-4M_{3}^{2}}$, and $z_{k}^{T}=z_{k}%
^{L}=0$, $z_{k}^{P}\simeq\left\vert \left(  \frac{2M_{1}}{M_{1}+M_{2}}\right)
^{\frac{1}{k}}\right\vert .$ The SSA in Eq.(\ref{tu}) reduces to%
\begin{align}
&  A_{st}^{(r_{n}^{T},r_{m}^{P},r_{l}^{L})}\nonumber\\
&  \simeq\prod_{n=1}\left[  (n-1)!\frac{\epsilon}{2}\sin\phi\right]
^{r_{n}^{T}}\prod_{m=1}\left[  -(m-1)!\frac{M_{1}+M_{2}}{2}\right]
^{r_{m}^{P}}\nonumber\\
&  \cdot\prod_{l=1}\left[  (l-1)!\frac{\epsilon}{2}\cos\phi\right]
^{r_{l}^{L}}B\left(  \frac{M_{1}M_{2}}{2},1-M_{1}M_{2}\right) \nonumber\\
&  \cdot F_{D}^{(K)}\left(  \frac{M_{1}M_{2}}{2};R_{m}^{P};M_{1}M_{2};\left(
\frac{2M_{1}}{M_{1}+M_{2}}\right)  _{m}\right)  ,
\end{align}
where $K=\sum_{j=1}^{m}j$. Note that for string states with $r_{k}^{P}=0$ for
all $k\geq2$, one has $K=1$ and the Lauricella functions in the low energy
nonrelativistic SSA reduce to the Gauss hypergeometry functions $F_{D}^{(1)}=$
$_{2}F_{1}.$ In particular, for the case of $r_{1}^{T}=N_{1}$, $r_{1}%
^{P}=N_{3}$, $r_{1}^{L}=N_{2}$, and $r_{k}^{X}=0$ for all $k\geq2$, the SSA
reduces to%
\begin{align}
&  A_{st}^{(N_{1},N_{2},N_{3})}\nonumber\\
=  &  \left(  \frac{\epsilon}{2}\sin\phi\right)  ^{N_{1}}\left(
\frac{\epsilon}{2}\cos\phi\right)  ^{N_{2}}\nonumber\\
\cdot &  \left(  -\frac{M_{1}+M_{2}}{2}\right)  ^{N_{3}}B\left(  \frac
{M_{1}M_{2}}{2},1-M_{1}M_{2}\right) \nonumber\\
\cdot &  _{2}F_{1}\left(  \frac{M_{1}M_{2}}{2};-N_{3};M_{1}M_{2};\frac{2M_{1}%
}{M_{1}+M_{2}}\right)  , \label{low}%
\end{align}
which agrees with the result obtained in \cite{LLY} previously.

\noindent\textit{Exact} \textit{symmetry of string scattering amplitudes} In
the Lie group approach of special functions, the associated Lie group for the
Lauricella function $F_{D}^{(K)}$ in the SSA at each fixed $K$ is the
$SL\left(  K+3,%
\mathbb{C}
\right)  $\ group \cite{sl4c} which contains the $SL\left(  2,%
\mathbb{C}
\right)  $\ fundamental representation of the $3+1$ dimensional spacetime
Lorentz group $SO(3,1)$. So $sl\left(  K+3,%
\mathbb{C}
\right)  $ contains the $2+1$ dimensional $so(2,1)$, the Lorentz spacetime
symmetry in our case as well. In the Regge limit, the Lauricella function in
the SSA reduces to the Appell function $F_{1}$ with associated group
$SL\left(  5,%
\mathbb{C}
\right)  $ \cite{sl5c}, which is $K$ independent. In the low energy
nonrelativistic limit, the Lauricella function in the SSA reduces to the Gauss
hypergeometry function $_{2}F_{1}$ with associated group $SL\left(  4,%
\mathbb{C}
\right)  $ \cite{sl4c}, which is also $K$ independent.

In sum, we have identified the \textit{exact} $SL\left(  K+3,%
\mathbb{C}
\right)  $ symmetry of string scattering amplitudes with three tachyons and
one\textit{ arbitrary} string state of $26D$ bosonic open string theory.
Finally, with the $SL\left(  K+3,%
\mathbb{C}
\right)  $\ group and the recurrence relations of the Lauricella functions
$F_{D}^{(K)}$, one can derive infinite number of recurrence relations of SSA
of different string states which are valid for \textit{all} energies. For a
simple example, the following recurrence relation of $F_{D}^{(K)}$ can be
verified%
\begin{align}
cF_{D}^{(K)}\left(  b_{j};c\right)  +c(x_{j}-1)F_{D}^{(K)}\left(
b_{j}+1;c\right)   &  \nonumber\\
+(a-c)x_{j}F_{D}^{(K)}\left(  b_{j}+1;c+1\right)   &  =0,\label{fd}%
\end{align}
which leads to the recurrence relation of SSA%
\begin{equation}
\left(  \frac{u}{2}+2-N\right)  A_{st}^{(r_{n}^{T},r_{m}^{P},r_{l}^{L}%
)}-\left(  \frac{s}{2}+1\right)  k_{3}^{T}A_{st}^{(r_{n}^{\prime T},r_{m}%
^{P},r_{l}^{L})}=0
\end{equation}
where $(r_{n}^{\prime T},r_{m}^{P},r_{l}^{L})$ means the group $\left(
-\{r_{1}^{T}-1\}^{1},\left\{  -r_{2}^{T}\right\}  ^{2},\cdots,\left\{
-r_{n}^{T}\right\}  ^{n};R_{m}^{P},R_{l}^{L}\right)  $ of polarizations. In
Eq.(\ref{fd}), we have omitted those arguments of $F_{D}^{(K)}$ which remain
the same for all three Lauricella functions:

\begin{acknowledgments}
This work is supported in part by the Ministry of Science and Technology and
S.T. Yau center of NCTU, Taiwan.
\end{acknowledgments}


\begin{thebibliography}{99}                                                                                               %


\bibitem {GM}D. J. Gross and P. F. Mende,
Phys.\ Lett.\ B \textbf{197}, 129 (1987);
Nucl.\ Phys.\ B \textbf{303}, 407 (1988).

\bibitem {Gross}D.~J.~Gross,
Phys.\ Rev.\ Lett.\ \textbf{60}, 1229 (1988); D. J. Gross and J. R. Ellis,
Phil.\ Trans.\ R. Soc. Lond. A329, 401 (1989).

\bibitem {GrossManes}D.~J.~Gross and J.~L.~Manes,
Nucl.\ Phys.\ B \textbf{326}, 73 (1989). See section 6 for details.

\bibitem {ChanLee1}C.~T.~Chan and J.~C.~Lee,
Phys.\ Lett.\ B \textbf{611}, 193 (2005).
J.~C.~Lee,
[arXiv:hep-th/0303012].

\bibitem {ChanLee2}C.~T.~Chan and J.~C.~Lee,
Nucl.\ Phys.\ B \textbf{690}, 3 (2004).


\bibitem {CHL}C.~T.~Chan, P.~M.~Ho and J.~C.~Lee,Nucl. Phys. B 708, 99 (2005).

\bibitem {PRL}C.~T.~Chan, P.~M.~Ho, J.~C.~Lee, S.~Teraguchi and Y.~Yang,
Phys. Rev. Lett. 96 (2006) 171601, \eprint{hep-th/0505035}.

\bibitem {CHLTY}C.~T.~Chan, P.~M.~Ho, J.~C.~Lee, S.~Teraguchi and Y.~Yang,
Nucl.\ Phys.\ B \textbf{725}, 352 (2005).


\bibitem {susy}C.~T.~Chan, J.~C.~Lee and Y.~Yang,
Nucl.\ Phys.\ B \textbf{738}, 93 (2006).


\bibitem {Moore}Gregory Moore, Finite in all directions. arXiv:hep-th/9305139, 1993.

\bibitem {Moore1}Gregory Moore, Symmetries of the bosonic string S-matrix. arXiv:hep-th/9310026,1993.

\bibitem {CKT}C.T. Chan, S. Kawamoto and D. Tomino, Nucl. Phys. B 885, 225 (2014).

\bibitem {review}J.C. Lee and Y. Yang, Review on High energy String Scattering
Amplitudes and Symmetries of String Theory, arXiv: 1510.03297.

\bibitem {KLY}S.L. Ko, J.C. Lee and Y. Yang, JHEP, 9060:028 (2009).

\bibitem {LY}J.C. Lee and Y. Mitsuka, JHEP 1304:082 (2013).

\bibitem {LY2014}J.C. Lee and Y. Yang, Phys. Lett. B739, 370 (2014).

\bibitem {LYY}J.C. Lee, C. H. Yan, and Y. Yang, "High energy string scattering
amplitudes and signless Stirling number identity", SIGMA, 8:045, (2012).

\bibitem {sl5c}Willard Miller. Jr., "Lie theory and the Appell functions
$F_{1}$", SIAM J. Math. Anal. Vol. 4 No. 4, 638 (1973).

\bibitem {LLY}S.H. Lai, J.C. Lee and Y. Yang, arXiv: 1601.0381.

\bibitem {sl4c}Willard Miller. Jr., "Lie theory and generalizations of the
hypergeometric functions", SIAM J. Appl. Math. Vol. 25 No. 2, 226 (1973).

\bibitem {BCJ}Z. Bern, J. J. M. Carrasco and H. Johansson, Phys. Rev. D 78,
085011 (2008) [hep-ph/0805.3993].

\bibitem {Appell}Joseph Kampe de Feriet and Paul Appell. Fonctions
hypergeometriques et hyperspheriques 1926.
\end{thebibliography}
\end{document}